\documentclass{article}
\usepackage{spconf}

\usepackage[utf8]{inputenc} 
\usepackage[T1]{fontenc}    
\usepackage{hyperref}       
\usepackage{url}            
\usepackage{booktabs}       
\usepackage{amsfonts}       
\usepackage{nicefrac}       
\usepackage{microtype}      
\usepackage{nccmath}        
\usepackage{ntheorem}       
\usepackage{cleveref}       
\usepackage{graphicx}       
\usepackage[table,xcdraw]{xcolor} 
\usepackage{amsmath, amssymb}  
\usepackage{bm}                
\usepackage{subcaption}        
\usepackage{hyperref}          
\usepackage{tikz}   
\usetikzlibrary{calc,intersections,through,scopes,arrows,decorations,backgrounds,positioning,fit,petri,automata,shapes}
\tikzstyle{box} = [draw, rectangle, rounded corners, thick, node distance=4em,
text width=3em, text centered, minimum height=3em]
\tikzstyle{container} = [draw, rectangle, dashed, inner sep=2em]
\tikzstyle{line} = [->, thick, -latex']

\graphicspath{{./fig/}}

\title{Attacking Speaker Recognition with \\Deep Generative Models}
\name{Wilson Cai, Anish Doshi, Rafael Valle}
\address{UC Berkeley}

\begin{document}
\maketitle
\begin{abstract}
    In this paper we investigate the ability of generative adversarial networks
    (GANs) to synthesize spoofing attacks on modern speaker recognition systems.
    We first show that samples generated with SampleRNN and WaveNet are
    unable to fool a CNN-based speaker recognition system. We propose  a
    modification of the Wasserstein GAN objective function to make use of data
    that is real but not from the class being learned. Our semi-supervised
    learning method is able to perform both targeted and untargeted attacks,
    raising questions related to security in speaker authentication systems. 
\end{abstract}

\section{INTRODUCTION} \label{sec:introduction}
Speaker authentication systems are being deployed for security critical
applications in industries like banking, forensics, and home automation. Like
other domains, such industries have benefited from recent advancements in deep
learning that lead to improved accuracy and trainability of the speech
authentication systems.  Despite the improvement in the efficiency of these
systems, evidence shows that they can be susceptible to adversarial
attacks\cite{wu2015spoofing}, thus motivating a current focus on understanding
adversarial attacks (\cite{szegedy2013intriguing},
\cite{goodfellow2014explaining}), finding countermeasures to detect and deflect
them and designing systems that are provably correct with respect to
mathematically-specified requirements~\cite{seshia2016vai}.

Parallel to advancements in speech authentication, neural speech
\textit{generation} (the process of using deep neural networks to generate
speech) has also seen huge progress in recent years \cite{wang2017tacotron}.  The combination of these advancements begs a natural
question that has, to the best of our knowledge, not yet been answered:
\begin{center}
Are speech authentication systems robust \\to adversarial attacks by speech generative models?
\end{center}

Generative Adversarial Networks (GANs) are generative models that recently have
been used to produce incredibly authentic samples in a variety of fields. The
core idea of GANs, a minimax game played between a generator network and a
discriminator network, extends naturally to the field of speaker authentication
and spoofing. 

With regards to this question, we offer in this research the following contributions:
\begin{itemize}
\item We evaluate samples produced with SampleRNN and WaveNet in their ability to fool text-independent speaker recognizers.
\item We propose strategies for untargeted attacks using Generative Adversarial Networks.
\item We propose a semi-supervised approach for targeted attacks by modifying
    Wasserstein's GAN loss function.
\end{itemize}


%
\section{RELATED WORK}\label{sec:related_work}
Modern generative models are sophisticated enough to produce fake\footnote{We
use the term fake to refer to computer generated samples} speech samples that
can be indistinguishable from real human speech. In this section, we provide a
summary of some existing neural speech synthesis models and their architectures.

WaveNet~\cite{van2016wavenet} is a generative neural network that is trained
end-to-end to model quantized audio waveforms. The model is fully probabilistic
and autoregressive, using a stack of causal convolutional layers to condition
the predictive distribution for each audio sample on all previous ones. It has
produced impressive results for generation of speech audio conditioned on
speaker and text and has become a standard baseline for neural speech generative
models. 

SampleRNN~\cite{mehri2016samplernn} is another autoregressive architecture that
has been successfully used to generate both speech and music samples. SampleRNN
uses a hierarchical structure of deep RNNs to model dependencies in the sample
sequence. Each deep RNN operates at a different temporal resolution so as to
model both long term and short term dependencies. 

Recent work on deep learning architectures has also introduced the presence of
\textit{adversarial examples}: small perturbations to the original inputs,
normally imperceptible to humans, which nevertheless cause the architecture to
generate an incorrect or deliberately chosen output. In their brilliant papers,
~\cite{szegedy2013intriguing} and ~\cite{goodfellow2014explaining} analyze the
origin of adversarial attacks and describe simple and very efficient techniques
for creating such perturbations, such as the fast gradient sign method (FGSM). 

In the vision domain, ~\cite{sharif2016accessorize} describe a technique for
attacking facial recognition systems. Their attacks are physically realizable
and inconspicuous, allowing an attacker to impersonate another individual. In
the speech domain,~\cite{carlini2016hidden} describe attacks on
speech-recognition systems which use sounds that are hard to recognize by humans
but interpreted as specific commands by speech-recognition systems.

To the best of our knowledge, GANs have not been used for the purpose of speech
synthesis\footnote{More specifically, Mel-Spectrogram synthesis}.
\cite{pascual2017segan} uses a conditional GAN for the purpose of speech
\textit{enhancement}, i.e. taking as input a raw speech signal and outputting a
denoised waveform. The model in \cite{chang2017learning} tackles the reverse
problem of using GANs to learn certain representations given a speech
spectrogram.

\section{ATTACKING SPEAKER RECOGNITION MODELS}\label{sec:spk_rec_atks}
\subsection{Neural speaker recognition system}
\label{sub:speaker_recognition}
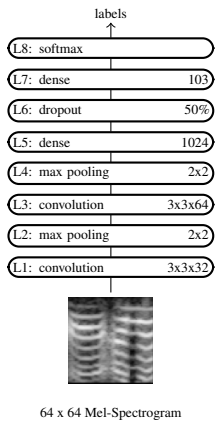
\begin{figure}[!ht]
    \begin{centering}
    \tikzset{ordinary/.style = {rectangle,draw,thick,rounded corners, minimum
        height = 0.5cm, minimum width=5cm, text width=5cm]}}
    \begin{tikzpicture}[node distance=0.8cm, scale=0.52, every
        node/.style={scale=0.52}]
       \node [] at (0,0) (start) {64 x 64 Mel-Spectrogram};
       \node [above=0.2cm of start] (a) {\includegraphics[width=.25\columnwidth]{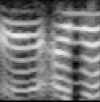}};
       \node [ordinary,above=0.2cm of a] (b) {L1: convolution \hfill 3x3x32};
       \node [ordinary,above of=b] (c) {L2: max pooling \hfill 2x2};
       \node [ordinary,above of=c] (d) {L3: convolution \hfill 3x3x64};
       \node [ordinary,above of=d] (e) {L4: max pooling \hfill 2x2};
       \node [ordinary,above of=e] (f) {L5: dense \hfill 1024};
       \node [ordinary,above of=f] (g) {L6: dropout \hfill 50\%};
       \node [ordinary,above of=g] (h) {L7: dense \hfill 103};
       \node [ordinary,above of=h] (i) {L8: softmax};
       \node [above=0.2cm of i] (j) {labels};
       \draw[-] (a) -- coordinate (a) (b);
       \draw[-] (b) -- coordinate (b) (c);
       \draw[-] (c) -- coordinate (c) (d);
       \draw[-] (d) -- coordinate (d) (e);
       \draw[-] (e) -- coordinate (e) (f);
       \draw[-] (f) -- coordinate (f) (g);
       \draw[-] (g) -- coordinate (g) (h);
       \draw[-] (h) -- coordinate (h) (i);
       \draw[->] (i) -- coordinate (i) (j);
    \end{tikzpicture}
    \caption{Architecture for CNN speaker verifier.}
    \label{fig:CNN}
    \end{centering}
\end{figure}

The speaker recognition system used in our experiments is based on the
framework by \cite{lukic2016speaker} and is described in Figure
\ref{fig:CNN}. The first module at the bottom is a pre-processing step that
extracts the Mel-Spectrogram from the waveform as described in section
\ref{sub:processdata}. The second module is a convolutional neural network (CNN)
that performs multi-speaker classification using the Mel-Spectrogram. The CNN is
a modified version of Alexnet~\cite{krizhevsky2012imagenet}. We warn the readers
that unlike~\cite{lukic2016speaker}, our classifier operates on 64 by 64 Mel-Spectrogram
and has slightly different number of nodes on each layer.

We train our speaker classifier using 64 by 64 Mel-Spectrograms~\footnote{64 mel
bands and 64 frames, 100 ms each} from 3 speech datasets, including 100 speakers
from NIST 2004, speaker p280 from CSTR VCTK and the single speaker in Blizzard.
Our speaker classifier has a rejection path, the “other” class, trained on
environmental sounds using samples from the ESC-50 dataset. Our model achieves
approximately $85\%$ test set accuracy

\subsection{Adversarial attacks}
We define adversarial attacks on speaker recognition systems as
\textit{targeted} or \textit{untargeted}. In targeted attacks, an adversary is
interested in designing an input that makes the classification system predict a
target class chosen by the adversary. In untargeted attacks, the adversary is
interested in a confident prediction, regardless of the class being predicted as
long as it is not the "other" class.  Untargeted attacks are essentially designed
to fool the classifier into thinking a fake speech sample is real. Note that a
successful targeted attack is by definition a successful untargeted attack as
well.

\section{EXPERIMENTAL SETUP}\label{sec:experiments}
\subsection{Datasets}
In our experiments we use three speech datasets and one dataset with
environmental sounds, as shown  in Table~\ref{tbl:datasets}. The datasets used
are public and provide audio clips of different lengths, quality, language and
content. In addition to the samples listed in Table~\ref{tbl:datasets}, we used
globally conditioned sampleRNN and WaveNet fake samples available on the web.
The samples generated with sampleRNN and WaveNet are from the Blizzard dataset
and CSTR VCTK (P280) respectively.
\begin{table}[!t]
\resizebox{\columnwidth}{!}{
\centering
\begin{tabular}{lllll}
                                                                     & \cellcolor[HTML]{C0C0C0}Speakers & \cellcolor[HTML]{C0C0C0}Language & \cellcolor[HTML]{C0C0C0}Duration & \cellcolor[HTML]{C0C0C0}Context \\ \cline{2-5} 
\multicolumn{1}{l|}{\cellcolor[HTML]{C0C0C0}2013 Blizzard} & 1                                & English                          & 73 h                             & Book narration                  \\
\multicolumn{1}{l|}{\cellcolor[HTML]{C0C0C0}CSTR VCTK}               & 109                    & English                          & 400 Sentences                    & Newspaper narration             \\
\multicolumn{1}{l|}{\cellcolor[HTML]{C0C0C0}2004 NIST}               & 100                    & Multiple                         & 5 min / speaker                  & Conversational phone speech.    \\                     
\multicolumn{1}{l|}{\cellcolor[HTML]{C0C0C0}ESC 50}                  & 50                     & N/A                             & 4 min / class                    & Environmental sounds.                          
\end{tabular}
}
\bigskip
\caption{Description of the datasets used in our experiments. }
\label{tbl:datasets}
\end{table}

\subsection{Pre-processing}
\label{sub:processdata}
Data pre-processing is dependent on the model being trained. For SampleRNN and
WaveNet, the raw audio is reduced to 16kHz and quantized using the $\mu$-law
companding transformation as referenced in~\cite{mehri2016samplernn}
and~\cite{van2016wavenet}. For the model based on the Wasserstein GAN,
we pre-process the data by converting it to 16kHz and removing silences by using
the WebRTC Voice Activity Detector (VAD) as referenced
in~\cite{zeidan2014webrtc}. For the CNN speaker recognition system, the data is
pre-processed by resampling to 16kHz when necessary and removing silences by
using the aforemetioned VAD. 

\subsection{Feature extraction}
SampleRNN and WaveNet operate at the sample level, i.e. waveform, thus requiring
no feature extraction. The features used for the neural speaker recognition
system are based on Mel-Spectrograms with dynamic range compression. The
Mel-Spectrogram is obtained by projecting a spectrogram onto a mel scale. We use
the python library librosa to project the spectrogram
onto 64 mel bands, with window size equal to 1024 samples and hop size equal to
160 samples, i.e. 100ms long frames. Dynamic range compression is computed as
described in~\cite{lukic2016speaker}, with $log(1 + C*M)$, where $C$ is a
compression constant scalar set to $1000$ and $M$ is a matrix representing the
Mel-Spectrogram. Training the GAN is also done with Mel-Spectrograms of 64 bands and 64 frames image patch.
                        
\subsection{Models}
\subsubsection{WaveNet}
Due to constraints on computing power and the extreme difficulty in training
WaveNet~\footnote{Our community has not been able to replicate the results in
Google's WaveNet paper}, we used samples from WaveNet models that had been
pre-trained for 88 thousand iterations. Parameters of the models were kept the
same as those in \cite{van2016wavenet}. \\ The ability of WaveNet to perform
\textit{untargeted} attacks amounts to using a model trained on an entire
corpus. Targeted attacks are more difficult - we found that a single speaker's
data was not enough to train WaveNet to converge successfully. To construct
speaker-dependent samples, we relied on samples from pre-trained models that
were \textit{globally conditioned} on speaker ID. Based on informal listening
experiments, such samples do sound very similar to the real speech of the
speaker in question.  

\subsubsection{sampleRNN}
Similarly to WaveNet, we found that the best (least noisy) sampleRNN samples
came from models which were pretrained with a high number of iterations.
Accordingly, we obtained samples from the three-tiered architecture, trained on
the Blizzard 2013 dataset \cite{prahallad2013blizzard}, which as mentioned in
Section 3 is a 300 hour corpus of a single female speaker's narration. We also
downloaded samples from online repositories, including samples from the original
paper's online repository at \texttt{https://soundcloud.com/samplernn/sets},
which we qualitatively found to have less noise than ours. 

\subsubsection{WGAN}
In all of our experiments, we use the Wasserstein GAN with gradient penalty
(WGAN-GP), which we found makes the model converge better than regular
WGAN~\cite{arjovsky2017wasserstein} or GAN~\cite{goodfellow2014generative}. 
In our experiments, we trained a WGAN-GP to produce mel-spectrograms from 1
target speaker \textit{against} a set of 101 speakers. On each critic iteration, we fed
it with a batch of samples from one target speaker, and a batch of data
uniformly sampled from the other speakers. We used two popular architectures
for generator/critic pairs: \textit{DCGAN}~\cite{radford2015unsupervised} and
\textit{ResNet}~\cite{ledig2016photo}. 

Performing \textit{untargeted} attacks with the WGAN-GP (i.e., training the
network to output speech samples that mimic the distribution of speech) is
relatively straightforward: we simply train the WGAN-GP using all speakers in
our dataset. However, the most natural attack is one that is \textit{targeted}:
where the GAN is trained to directly fool a speaker recognition system, i.e., to
produce samples that the system classifies as matching a target speaker with
reasonable confidence.

\subsubsection{WGAN-GP with modified objective function}
A naive approach for targeted attacks is to train the GAN on the data of the
single target speaker. A drawback of this approach is that the \textit{critic},
and by consequence the \textit{generator}, does not have access to universal
properties of speech\footnote{We draw a parallel with Universal Background
Models in speech.}. To circumvent this problem, we rely on semi-supervised
learning and propose a modification to the critic's objective function that
allows it to learn to differentiate between not only real samples and generated
samples, but also between real speech samples from a target speaker and real
speech samples from other speakers. We do this by adding a term to the critic's
loss that encourages its discriminator to classify real speech samples from
untargeted speakers as fake: 
\normalsize

\tiny
\begin{align}
\medmath{
    \underbrace{\underset{\boldsymbol{\widetilde{x}} \sim
    \mathbb{P}_{g}}{\mathbb{E}}
    \big[D(\boldsymbol{\widetilde{x}})\big]}_\text{Generated Samples}
    \color{red} +  \underbrace{\alpha * \underset{\boldsymbol{\dot{x}} \sim
    \mathbb{P}_{\dot{x}}}{\mathbb{E}}
    \big[D(\boldsymbol{\dot{x}})\big]}_\text{Different Speakers} \color{black} -
    \underbrace{\underset{\boldsymbol{x} \sim \mathbb{P}_{r}}{\mathbb{E}}
    \big[D(\boldsymbol{x})\big]}_\text{Real Speaker}  + \underbrace{\lambda
    \underset{\boldsymbol{\hat{x}} \sim \mathbb{P}_{\hat{x}}}{\mathbb{E}}
    \big[(\lVert \nabla_{\boldsymbol{\hat{x}}} D(\boldsymbol{\hat{x}}) \rVert_2
    - 1)^2\big]}_\text{Gradient Penalty}\label{eq:wgan_gp_mixed},
}
\end{align}
\normalsize
where $P_{\hat{x}}$ is the distribution of samples from other speakers and
$\mathbf{\alpha}$ is a tunable scaling factor. Note that
equation~\ref{eq:wgan_gp_mixed} is no longer a direct approximation of the
Wasserstein distance. Rather, it provides a balance of the distance between both
the fake distribution and real one, and the distance between other speakers'
distribution and the target speaker's one. We refer to this objective function
as \textbf{mixed loss}.

Initially, we were able to converge the targeted loss model used the same
parameters as \cite{gulrajani2017improved}, namely 5 critic iterations per
generator iteration, a gradient penalty weight of 10, and batch size of 64. Both
the generator and critic were trained using the Adam optimizer
\cite{kingma2014adam}. However, under these parameters we found that the highest
$\alpha$ weight we could successfully use was 0.1 (we found that not including
this scaling factor led to serious overfitting and poor convergence of the GAN).

In order to circumvent these problems and train a model with $\alpha$ set to 1,
we made modifications to the setup, including setting the standard deviation of
the DCGAN discriminator's weight initialization to 0.05 and iterations to 20. To
accommodate the critic's access to additional data in the mixed loss function
(4), we increased the generator's learning rate. Finally, we added Gaussian noise
to the target speaker data to prevent overfitting. 

\section{RESULTS}\label{sec:results}
\subsection{GAN Mel-Spectrogram}
Using the improved Wasserstein GANs framework, we trained generators to
construct 64x64 mel-spectrogram images from a noise vector. Visual results are demonstrated below in Figure~\ref{fig:samples_comparison}.   We saw recognizable Mel-Spectrogram-like features in the
data after only 1000 generator iterations, and after 5000 iterations the generated samples were indistinguishable from real ones. Training took around 10
hours for 20000 iterations on a single 4 GB Nvidia GK104GL GPU.
\begin{figure}[!t]
    \centering
    \begin{subfigure}[t]{0.3\columnwidth}
        \centering
        \includegraphics[width=\columnwidth]{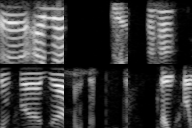}
        \caption{Real (actual)}
        \label{fig:samples_real}
    \end{subfigure}
    \qquad
    \begin{subfigure}[t]{0.3\columnwidth}
        \centering
        \includegraphics[width=\columnwidth]{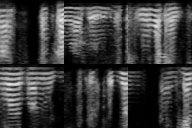}
        \caption{Fake (generated)}
        \label{fig:samples_fake}
    \end{subfigure}
    \caption{Comparison of 6 real and fake mel-spectrogram samples from all speakers. ($\sim$ 5000 generator iterations) }
    \label{fig:samples_comparison}
\end{figure}

\subsection{GAN Adversarial attacks}

Within the GAN framework, we train models for untargeted attacks by using all
data available from speakers that the speaker recognition systems was trained
on, irrespective of class label. We show in subsection \ref{sub:untargeted} that
an untargeted model able to generate data from the real distribution with enough
variety can be used to perform adversarial attacks.
Figure~\ref{fig:conf_mat_untargeted} depicts that our GAN-trained generator
successfully learns all speakers across the dataset, without mode collapsing.

As we described earlier, the models for targeted attacks can be trained in two
manners: 1) conditioning the model on additional information, e.g. class labels,
as described in~\cite{mirza2014conditional}; 2) using only data from the label
of interest. While the first approach might result in mode collapse, a drawback
of the second approach is that the discriminator, and by consequence the
generator, does not have access to universal\footnote{We draw a parallel with
Universal Background Models in speech.} properties of speech. In the targeted
attacks subsection \ref{sub:targeted} we show results using our new objective
function described in equation~\ref{eq:wgan_gp_mixed} that allows using data
from all speakers.  

\subsubsection{Untargeted attacks}\label{sub:untargeted}
For each speaker audio data in the test set, we compute a Mel-Spectrogram as
descibred in section \ref{sub:processdata}. The resulting Mel-Spectrogram is
then fed into the CNN recognizer and we extract a 1024-dimensional feature $\Phi$ from
the first fully-connected layer (L5) in the pre-trained CNN model
(\ref{fig:CNN}) trained on the real speech dataset with all speaker IDs. This
deep feature/embedding $\Phi$ is then used to train a K-nearest-neighbor (KNN)
classifier, with K equal to 5.

To control the generator trained by our WGAN-GP, we feed the generated
Mel-Spectrograms into the same CNN-L7 pipeline to extract their corresponding
feature $\widehat \Phi$. Utilizing the pre-trained KNN, each sample is assigned to
the nearest speaker in the deep feature space. Therefore, we know which speaker
our generated sample belongs to when we attack our CNN recognizer. We evaluate our
controlled WGAN-GP samples against our CNN speaker recognition system, and the
confusion matrix can be found in Figure \ref{fig:conf_mat_untargeted}. 

\subsubsection{Targeted attacks}\label{sub:targeted}
\begin{figure}[!t]
    \centering
    \begin{subfigure}[b]{0.4\columnwidth}
        \centering
        \includegraphics[width=\columnwidth]{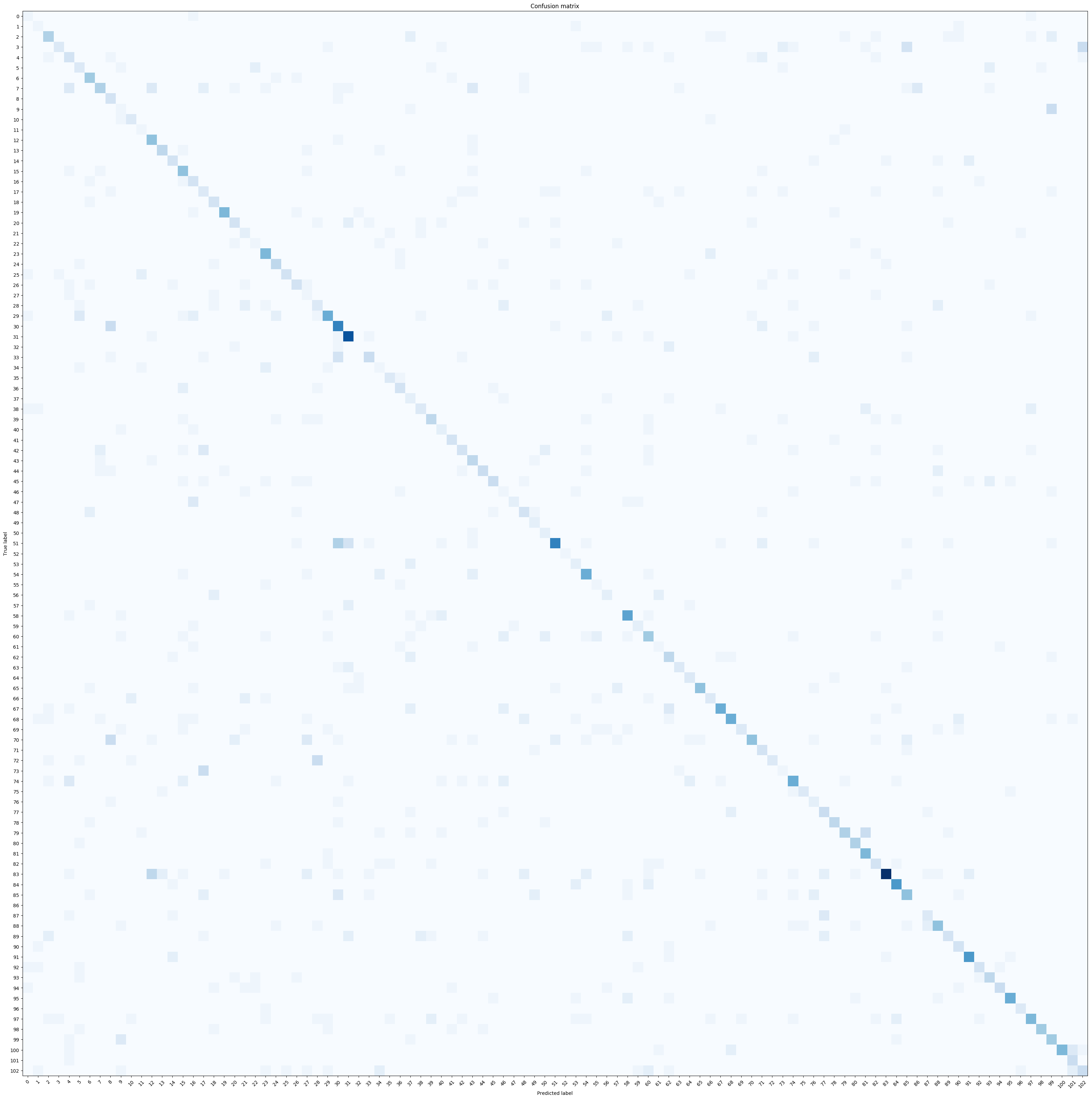}
        \caption{Confusion matrix of untargeted model. x-axis corresponds to predicted label, y-axis to ground truth.}
        \label{fig:conf_mat_untargeted}
    \end{subfigure}  
    \qquad
    \begin{subfigure}[b]{0.5\columnwidth}
        \centering
        \includegraphics[width=\columnwidth]{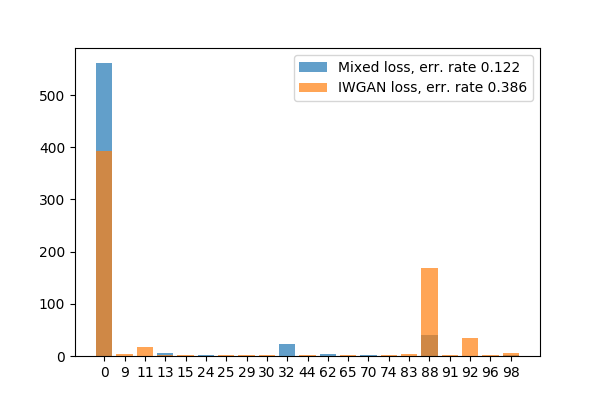}
        \caption{Histogram of predictions given WGAN-GP and mixed loss models. Ground truth label: 0.}
        \label{fig:pred_comp_spk0}
    \end{subfigure}
    \caption{Summary histograms of targeted attacks}
    \label{fig:confusion_matrices}
\end{figure}

We trained the WGAN-GP on the entirety of the NIST 2004 corpus (100 speakers), a single speaker (P280) from
the VCTK Corpus, and the single speaker from the Blizzard dataset. The samples
from the other models were either downloaded from the
web or created from WaveNet globally conditioned on the single VCTK corpus
speaker, and on SampleRNN trained only on data from the Blizzard dataset.
Results for the WGAN-GP are demonstrated in Figure \ref{fig:confusion_matrices}.  
In the samples generated with sampleRNN and WaveNet models, \textbf{none} of the
predictions made by the classifier match the target speaker. 

We also trained the WGAN-GP with and without the \textbf{mixed loss} on
different speakers. The histogram of predictions in Figure
~\ref{fig:pred_comp_spk0} shows WGAN-GP results for speaker 0. The improved
WGAN-GP loss achieves 0.38 error rate and our mixed loss achieves 0.12 error
rate, producing a 75\% increase in accuracy. 

\section{DISCUSSION AND CONCLUSION}\label{sec:conclusions}
In this research we have investigated the use of speech generative models to
perform adversarial attacks on speaker recognition systems. We show that the
samples from autoregressive models we trained, i.e. SampleRNN and WaveNet, or
downloaded from the web were not able to fool the CNN speaker recognizers we
used in this research. On the other hand, we show that adversarial examples
generated with GAN networks are successful in performing targeted and untargeted
adversarial attacks given the speaker recognition used herein.

\vfill\pagebreak

\bibliographystyle{IEEEbib}
\bibliography{references}

\end{document}